\documentclass{ws-rv961x669}
\usepackage{ws-rv-van}     
\usepackage{ws-rv-thm}     
\usepackage{subfigure}     
\makeindex

\newcommand{\beq}{\begin{equation}}
\newcommand{\eeq}{\end{equation}}
\newcommand{\beqa}{\begin{eqnarray}}
\newcommand{\eeqa}{\end{eqnarray}}

\begin{document}

\chapter[Photon Scattering off Nuclei]{Photon Scattering off Nuclei$^*$}\footnote{Dedicated
to the  memeory of Walter Greiner.}

\author[Hartmuth Arenh\"ovel]{Hartmuth Arenh\"ovel}

\address{Institut f\"{u}r Kernphysik, Johannes Gutenberg-Universit\"{a}t, \\
   D-55099 Mainz, Germany }

\begin{abstract}
The study of nuclear and subnuclear structure by means of photon
scattering is outlined. Besides a brief exposition of the formalism a
few illustrative examples are discussed. 
\end{abstract}

\section{Introduction}

I would like to begin with a brief personal remark: It was during the fall of
1964 - I just had completed my diploma thesis in experimental physics
at the University of Freiburg - , when Walter Greiner offered me to
work on a PhD thesis in his new established theory group at the
University of Frankfurt. As subject of the thesis he had proposed to
investigate the structure of heavy deformed nuclei in the region of
the giant resonances by means of photon scattering. By the end of 1965
Hans-J\"urgen Weber and myself were Walter's first PhD-students to
complete their PhD. Since then, my interest in the study of
electromagnetic reactions on nuclei in general, e.g.\ photo
absorption and scattering, electron scattering and meson production,
has continued up to present times. 

Electromagnetic reactions on nuclei provide an excellent
tool to investigate nuclear structure. In addition, they also lead
to valuable insights into the electromagnetic properties of the
nuclear constituents, proton and neutron, like for example, electric
and magnetic polarizabilities and electromagnetic form
factors. In this context photon scattering experiments are a
particularly interesting source of information on off-shell
properties of the nuclear constituents. On the other hand genuine
microscopic calculations of 
photon scattering cross sections are rather complicated since in
principle the complete excitation spectrum has to be taken into
account. 

Here I will give a brief account of photon scattering reactions on
nuclei with emphasis on my own work and that of my collaborators. It is
not intended as a general review, rather a personal view on this
interesting reaction which over many years has fascinated myself and
which was and still is subject of my own research.  Early reviews on
nuclear photon scattering may be found in
Refs.~\refcite{ArG69,Are84,Are86} and a more recent one in 
Ref.~\refcite{HLMS00}.  

In the next section I will present a short summary of the 
basic scattering formalism, in particular the expansion of the
scattering matrix in terms of generalized polarizabilities as
basic quantities. It will be followed by a few 
illustrative examples, partly on earlier work on medium and heavy
weight nuclei within the dynamic collective model of the giant dipole
resonances and partly on the lightest nucleus, the deuteron, with
emphasis on subnucleon degrees of freedom like meson exchange and
isobar currents.  

\section{Formalism of photon scattering}

I will briefly describe the formal features of the photon scattering process 
\beq
\gamma_\lambda (k) + N_i(P_{i} ) \longrightarrow
\gamma'_{\lambda'}(k') + N_f(P_{f}) \,, 
\eeq
where an incoming photon with four momentum $k=(k_0, \vec k)$
and circular polarization $\vec e_{\lambda}$
($\lambda=\pm 1$) is scattered off
a nucleus in the initial intrinsic state $|i\rangle$ with total four momentum
$P_{i}=(E_i,\vec P_i)$ making a transition to a
final intrinsic state $|f\rangle$ with total four momentum 
$P_{f}=(E_f,\vec P_f)$ while emitting a final
photon  with four momentum $k'=(k_0',\vec k^{\,\prime})$ and
polarization $\vec  e_{\lambda'}^{\,\prime}$.

\subsection{The photon scattering amplitude}

\begin{figure}
\centerline{\includegraphics[width=9cm]{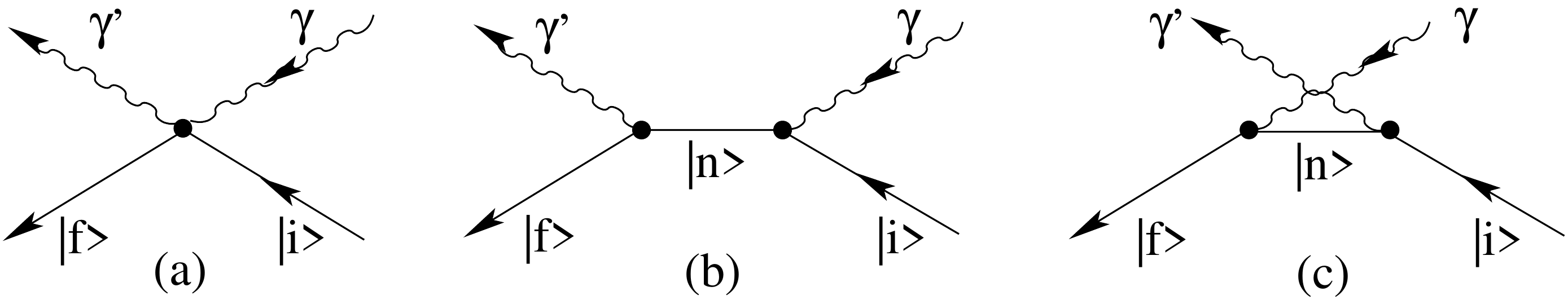}}  
\caption{Diagrammatic representation of photon scattering: The
  two-photon amplitude  (a) and the resonance contribution (direct (b) and
  crossed (c)). To be read from right to left.} 
\label{Fig-scattering-diagram} 
\end{figure}

Since the electromagnetic interaction is weak one makes a Taylor
expansion with respect to the electromagnetic field $A_\mu$ up to second
order, because at least two photons are involved in the scattering
process. Thus in this lowest,
i.e.~second order in the e.m.~coupling, the scattering amplitude is
given by two terms, the contact or two photon amplitude (TPA or
seagull) $B _{\lambda'\lambda}(\vec k^{\,\prime},\vec k)$, arising
from the second order term of the Taylor expansion, and the 
resonance amplitude (RA) $R _{\lambda'\lambda}(\vec k^{\,\prime},\vec
k)$ from the iterated linear interaction term. A diagrammatic 
illustration is shown in Fig.~\ref{Fig-scattering-diagram}. 

Accordingly, the total scattering amplitude is the sum of these two
contributions 
\beq
T^{fi}_{\lambda'\lambda}(\vec k^{\,\prime},\vec k)=
B^{fi}_{\lambda'\lambda}(\vec k^{\,\prime},\vec k)
+R^{fi}_{\lambda'\lambda}(\vec k^{\,\prime},\vec k)\,,\label{scattering-amplitude}
\eeq
where the two-photon amplitude (diagram (a) of
Fig.~\ref{Fig-scattering-diagram}) has the form 
\beq
B^{fi}_{\lambda'\lambda}(\vec k',\vec k)=\sum_{l,m=1}^3
e_{\lambda',l}^{\,\prime *} \langle f| \widehat B_{lm}(\vec k^{\,\prime},\vec
k\,) |i\rangle e_{\lambda,m}\,,
\eeq
with $\widehat B_{lm}(\vec k^{\,\prime},\vec k\,)$ as Fourier
transform of the second order coefficient of the Taylor expansion
$\widetilde B_{lm}(\vec x,\vec y\,)$, i.e.
\beq
\widehat B_{lm}(\vec k^{\,\prime},\vec k\,) = -\int d^3x\, d^3y\,
e^{i\vec k^{\,\prime}\cdot\vec x}
\widetilde B_{lm}(\vec x,\vec y\,) e^{-i\vec k\cdot\vec y}\,.
\label{TP-amplitude}
\eeq
Assuming a nonrelativistic description, the overall
center-of-mass (c.m.) motion can be separated and the Hamiltonian
splits accordingly into an intrinsic part and the c.m.\ kinetic energy
\beq
H=H_{int}+\frac{\vec P^{\,2}}{2M_A}\,,
\eeq
where $H_{int}=T_{int} +V$ denotes the Hamiltonian of the internal
motion, $\vec P$ the total c.m.\ momentum and $M_A$ the mass of
the nucleus. An intrinsic state is denoted by
$|n\rangle$ with intrinsic energy $e_n$.
Then the resonance amplitude (RA) (diagrams (b) and (c)
of Fig.~\ref{Fig-scattering-diagram})  is given by the following
matrix element between intrinsic states
\beqa
R^{fi}_{\lambda'\lambda}(\vec k^{\,\prime},\vec k)=
\langle f|&\Big[&\vec e_{\lambda'}^{\,\prime *}\cdot \vec 
  J(-\vec k^{\,\prime},2\vec P_f+\vec k^{\,\prime}) \,
\nonumber\\&&\times
G_{int}\Big(k_0-\frac{\vec k \cdot(2\vec P_i+\vec k)}{2M_A}
+i\epsilon\Big)\,
\vec e_{\lambda}\cdot \vec 
  J(\vec k,2\vec P_i+\vec k)
\nonumber\\ && 
+\left(
\vec e_{\lambda} \leftrightarrow 
\vec e_{\lambda'}^{\,\prime *}\,,
k_\mu \leftrightarrow -\vec k^{\,\prime}_\mu
\right)
\Big]|i\rangle
\,,\label{R-amplitude}
\eeqa
with the resolvent or propagator
\beq
G_{int}(z)=(H_{int}-e_i -z)^{-1}\,.
\eeq
The current operator in Eq.~(\ref{R-amplitude})
\beq
\vec J(\vec k,\vec P)=\vec j(\vec k)
+\frac{1}{2M_A}\,\vec P\,\rho(\vec k)\label{current}
\eeq
consists of the intrinsic current density operator $\vec j(\vec k\,)$
plus a term taking into account the convection current of the
separated c.m.~motion, where $\rho(\vec k\,)$ denotes the Fourier
transform of the intrinsic charge density operator of the nucleus. 
The intrinsic charge and current density operators
consist of a kinetic or one-body ($\rho_{[1]},\vec j_{[1]}$) and a
two-body meson exchange part ($\rho_{[2]},\vec j_{[2]}$)
\beqa
\rho(\vec k\,)&=&\rho_{[1]}(\vec k\,)+\rho_{[2]}(\vec k\,)\,,\\
\vec j(\vec k\,)&=&\vec j_{[1]}(\vec k\,)+\vec j_{[2]}(\vec k\,)\,,
\eeqa
with
\beqa
\rho_{[1]}(\vec k)&=&\sum_le_l\,e^{-i\vec k\cdot \vec r_l}\,,\\
\vec j_{[1]}(\vec k\,)&=&\frac{1}{2M}\sum_l \Big(e_l \{\vec p_l,e^{-i\vec
  k\cdot \vec r_l}\} + \mu_l \vec \sigma_l\times\vec k\,e^{-i\vec
  k\cdot \vec r_l}\Big)\,.
\eeqa
Here, $e_l$ and $\mu_l$ denote charge and magnetic moment of the
$l$-th particle and $\vec r_l$, $\vec p_l$ and $\vec \sigma_l$ its internal
coordinate, momentum and spin operators, respectively. 
The expressions for the corresponding exchange operators depend on the
interaction model. At least in the nonrelativistic limit, the exchange
contribution to the charge density vanishes (Siegert's hypothesis). 
Furthermore, also the TPA consists of a kinetic one-body contribution
and a two-body exchange amplitude 
\beq
 \widehat B_{lm}(\vec k^{\,\prime},\vec
k\,)= \widehat B_{[1],lm}(\vec k^{\,\prime},\vec
k\,)+ \widehat B_{[2],lm}(\vec k^{\,\prime},\vec k\,) \,,
\eeq
where the kinetic one-body operator is given by the sum of the
individual proton Thomson scattering amplitudes
\beq
\widehat B _{[1],lm}(\vec k^{\,\prime},\vec
k\,)= -\frac{1}{M}\sum_j e_j^2 \,e^{-i(\vec k-\vec k^{\,\prime})
\cdot \vec r_j}\,\delta_{lm}\,.
\eeq

It is important to note that the splitting of the scattering amplitude
into a resonance and a 
two-photon amplitude is gauge dependent. This gauge dependence is
reflected in gauge conditions for the current and the two-photon
amplitude which follow from the Gauge invariance of the
electromagnetic interaction. In detail one finds the following gauge
conditions for the e.m.~operators 
\beqa
\vec k\cdot\vec j(\vec k\,)&=&[H,\rho(\vec k\,)]\,,\label{gauge1}\\
\sum_l k^{\,\prime}_l B_{lm}(\vec
k^{\,\prime},\vec k\,)&=& [\rho(-\vec k^{\,\prime}\,), j_m(\vec k\,)]\,.
\label{gauge2}
\eeqa
The first condition in (\ref{gauge1}), connecting the charge density
with the current, describes current conservation, while the second
relates the TPA to the commutator of charge and current
densities. Separating the one-body and two-body (exchange)
contributions, one finds the following conditions 
\beqa
\vec k\cdot\vec j_{\,[1]}(\vec k\,)&=&[T,\rho_{\,[1]}(\vec k\,)]\,,\\
\vec k\cdot\vec j_{\,[2]}(\vec k\,)&=&[V,\rho_{\,[1]}(\vec
k\,)]+[T,\rho_{\,[2]}(\vec k\,)]\,,\\ 
\sum_l k^{\,\prime}_l \widehat B_{[1],lm}(\vec
k^{\,\prime},\vec k\,)&=& [\rho_{\,[1]}(-\vec k^{\,\prime}\,), j_{[1],m}(\vec
k\,)]\,,\\ 
\sum_l k^{\,\prime}_l \widehat B_{[2],lm}(\vec
k^{\,\prime},\vec k\,)&=& [\rho_{\,[1]}(-\vec k^{\,\prime}\,), j_{\,[2],m}(\vec
k\,)]+[\rho_{\,[2]}(-\vec k^{\,\prime}\,), j_{\,[1],m}(\vec
k\,)]\,. 
\eeqa
Important consequences are the low energy limits~\cite{Fri75,ArW86}
\beqa
 \vec j(0)&=&[H,\vec D]\,,
\eeqa
the Siegert theorem, where $\vec D$ denotes the unretarded dipole
operator, and 
\beqa
\widehat B^{ii}_{[1],l m}(0,0)&=&
-\frac{Ze^2}{M}\delta_{l m}\,,\\
\widehat B^{ii}_{[2],
l m}(0,0)&=&
-\langle i|[D_l,[V,D_m]]|i\rangle \,,\\
\widehat R^{ii}_{l m}(0,0)&=&
\frac{NZe^2}{AM}\delta_{l m}-\widehat B^{ii}_{[2],l m}(0,0)\,,
\eeqa
resulting in the classical Thomson limit for the total nuclear
scattering amplitude 
\beqa
T^{ii}_{\lambda'\lambda}(0,0)&=&
-\vec e_{\lambda'}^{\,\prime *}\cdot \vec
e_{\lambda}\frac{(Ze)^2}{AM}\,,
\label{Thomson}
\eeqa
with the approximation $M_A=AM$.

To close this section, I would like to mention the optical theorem,
which relates the forward elastic scattering amplitude to the total photo
absorption cross section 
\beq
\sigma_{tot}(k,\rho)=\frac{4\pi}{k}\,
Im\Big[Tr\Big(\rho \,T^{ii}(\vec k,\vec k\,)\Big)\Big]\,,\label{opt-th}
\eeq
where $\rho$ denotes the photon-nucleus polarization density
matrix of the initial state.

\subsection{Generalized nuclear polarizabilities}

In the scattering process the incoming photon and the scattered one
transfer various angular momenta via the e.m.\ multipole operators $L$
and $L'$ to the nucleus which can be coupled to a total angular
momentum transfer $J$. It is very useful to
expand the scattering amplitude with respect to this total angular
momentum transfer. It leads to the concept of generalized
polarizabilities $P_{fi,\lambda' \lambda}^{(L'L) J}(k',k)$, first
introduced by Fano~\cite{Fan60} for the case 
of pure $E1$ transitions and later generalized in
Refs.~\refcite{Ar65,ADG67,SiU68,Sil68}. A review may be found in
Ref.~\refcite{ArG69}. A further generalization to the  $(e,e'\gamma)$
reaction of electron scattering (virtual compton  scattering) is given
in Ref.~\refcite{ArD74}.  

The polarizabilities allow one to separate geometrical aspects related
to the angular momentum properties from dynamical effects as contained
in the strength of the various polarizabilities. 
The expansion of the total scattering
amplitude in terms of these polarizabilities reads
\beqa
T^{fi}_{M_f M_i, \lambda'\lambda}(\vec k^{\,\prime},\vec k)&=&
(-)^{I_f-M_i}\sum_{L',M',L,M,J,m}(-)^{L+L'}{\widehat J}^2
\left(\begin{matrix}
I_f&J& I_i \cr -M_f &m&M_i \cr
\end{matrix}\right) \nonumber\\&&
\times \left(\begin{matrix}
L&L'& J \cr M &M'&-m \cr
\end{matrix}\right)
P_{fi,\lambda' \lambda}^{(L'L) J}(k',k)
D^L_{M,\lambda}(R)D^{L'}_{M',-\lambda'}(R')\,,\label{T-pol}
\eeqa
where the abbreviation $\widehat J=\sqrt{2J+1}$ is used, 
 and $(I_i,M_i)$ and $(I_f,M_f)$ refer to the angular momenta of the
initial and final states and their projections on the quantization axis,
respectively. Furthermore, $R$ and $R'$ describe the rotations which
carry the quantization axis into the directions of the photon momenta
$\vec k$ and $\vec k^{\,\prime}$, respectively, and
$D^L_{M,\lambda}(R)$ and $D^{L'}_{M',-\lambda'}(R')$ denote the corresponding
rotation matrices in the convention of Rose~\cite{Ros57}. The
polarizabilities contain the dynamic properties of the system and
depend on the absolute values of the photon momenta only, whereas the
geometrical aspects, i.e.\ the angular dependencies, are contained in
the rotation matrices. 
 
The general definition of the polarizabilies is then obtained by the
inversion of Eq.~(\ref{T-pol}), i.e.
\beqa
P_{fi,\lambda' \lambda}^{(L'L) J}(k',k)&=&\frac{(-)^{L'-L-I_f}}{8\pi^2}
\widehat{L'}^2 \widehat{L}^{\,2}\sum_{M_f M_i M' M m}
\left(\begin{matrix}
I_f&J& I_i \cr -M_f &m&M_i \cr
\end{matrix}\right) 
\left(\begin{matrix}
L&L'& J \cr M &M'&-m \cr
\end{matrix}\right) \nonumber\\&&
\times\int dR'\,\int dR\,
D^{L*}_{M,\lambda}(R)D^{L'*}_{M',-\lambda'}(R')
\,T^{fi}_{M_f M_i, \lambda'\lambda}(\vec k^{\,\prime},\vec k)\,.
\eeqa
One should note that $J$ is bound by the multipole orders $L$
and $L'$ and the spins $I_i$ and $I_f$ of the initial and final
nuclear states, respectively, i.e.
\beqa
|L-L'|\le J\le L+L'\quad\mathrm{and}\quad |I_i-I_f|\le J\le I_i+I_f\,.
\eeqa
For example, to elastic scattering off a spin-zero nucleus only the
scalar polarizabilities ($J=0$) contribute. 

It is furthermore useful to classify the polarizabilities according to
the total parity transfer in case that parity is conserved. This leads
to the introduction of
\beq
P_{fi}^{ J}(M^{\nu'}L',M^{\nu}L
,k',k)=\frac{1}{4}\sum_{\lambda',\lambda=\pm 1}
\lambda^{\prime \nu'}\lambda^{\nu}\,
P_{fi,\lambda' \lambda}^{(L'L) J}(k',k)\,.
\eeq
In terms of these one has
\beq
P_{fi,\lambda' \lambda}^{(L'L) J}(k',k)=\sum_{\nu',\nu=0,1}
\lambda^{\prime\,\nu'}\lambda^\nu 
P_{fi}^{ J}(M^{\nu'}L',M^{\nu}L ,k',k)\,,
\eeq 
where $\nu$ and $\nu'$ classify the type of multipole transition,
i.e.\ $\nu=0$ means electric ($M^{0}L =EL$) and $\nu=1$ magnetic
($M^{1}L=ML$).   

\begin{figure}[ht]
\centerline{\includegraphics[width=11cm]{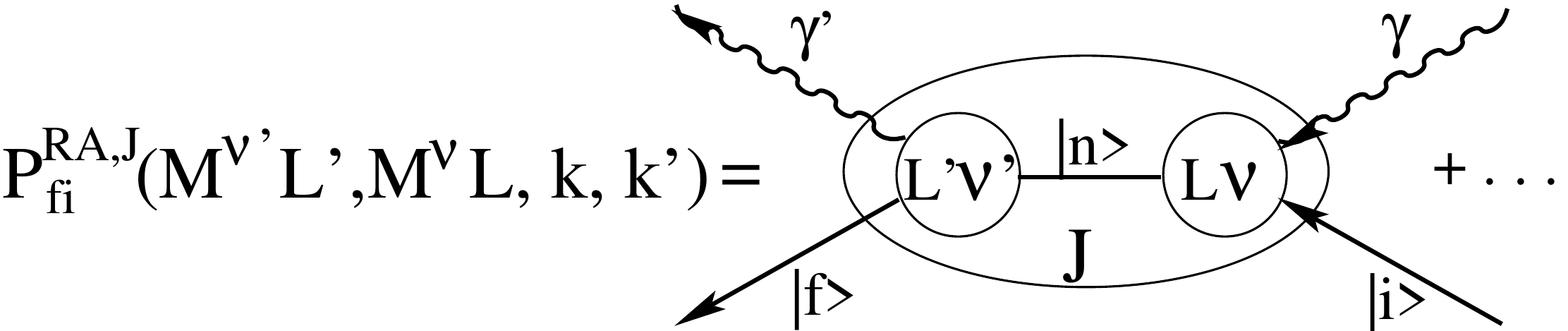}}
\caption{Graphical representation of the contribution of the direct
  term of the  resonance amplitude to the generalized polarizability
 $P_{fi}^{ J}(M^{\nu'}L',M^{\nu}L ,k',k) $, where the angular momentum 
 transfer $L$ of the incoming photon is coupled with the angular
 momentum transfer $L'$ of the outgoing photon to a total angular
 momentum transfer $J$.} 
\label{Fig-polarizability} 
\end{figure}

For parity conservation a simple selection rule follows
\beq
P_{fi}^{ J}(M^{\nu'}L',M^{\nu}L ,k',k)=0\,, \quad \mathrm{if}\quad
(-)^{L'+\nu'+L+\nu}\neq \pi_i\pi_f\,,
\eeq
with $\pi_i$ and $\pi_f$ denoting the parities of intial and final states,
respectively. A graphical visualization of 
the generalized polarizability is shown in
Fig.~\ref{Fig-polarizability} for the direct resonance term.

The polarizabilities $P_{fi}^{ J}(M^{\nu'}L',M^{\nu} L,k',k)$ can be separated
into a TPA  and a resonance contribution
\beqa
P_{fi}^{ J}(M^{\nu'}L',M^{\nu}L,k',k)&=&P_{fi}^{TPA,  J}(M^{\nu'}L',M^{\nu}L,k',k)
\nonumber\\&&+P_{fi}^{RA,  J}(M^{\nu'}L',M^\nu L,k',k)\,,
\eeqa 
where for the resonance amplitude one has
\beqa\label{compact-res}
P_{fi}^{RA, J}(M^{\nu'}L',M^{\nu}L,k',k)&=&2\pi(-)^{L+J}\frac{\hat L\hat
  L'}{\hat J}\nonumber\\
&&\hspace*{-3.3cm}\times\langle I_f e_f||\, \Bigg(\Big[M^{\nu'[L']}(k')
G_{int}(k_0-\frac{\vec k \cdot(2\vec P_i+\vec k)}{2M_A}+i\varepsilon)
M^{\nu [L]}(k)\Big]^{[J]}\nonumber\\
&&\hspace*{-1cm}
+\left\{
M^{\nu [L]}(k)\leftrightarrow M^{\nu'[L']}(k')\,,
k_\mu \leftrightarrow -\vec k^{\,\prime}_\mu
\right\}
\Bigg)||I_i e_i\rangle\,.
\eeqa
The superscript $[J]$ indicates a spherical tensor of rank $J$, and 
``$[\dots]^{[J]}$'' means that two spherical tensors 
are coupled to a spherical tensor of rank $J$. 
Furthermore, $M^{\nu' [L']}(k')$ denotes a standard electromagnetic
current multipole operator~\cite{Ros57}
\beqa
M^{\nu' [L']}_m(k')=\left\{\begin{matrix}
\nu=0 \quad\mathrm{electric}: & \int d^3x\, \vec A^{\,[L]}_m(E;k,\vec x\,)\cdot \vec J(\vec x\,)\,,\cr
\nu=1 \,\mathrm{magnetic}: & \int d^3x\, \vec A^{\,[L]}_m(M;k,\vec x\,)\cdot \vec J(\vec x\,)\,,
\end{matrix}
\right.
\eeqa
with the multipole fields 
\beqa
\vec A^{\,[L]}_m(M;k,\vec x\,)&=&i^Lj_L(kx)\vec Y^{\,(L1)[L]}_m(\hat x)\,,\\
\vec A^{\,[L]}_m(E;k,\vec x\,)&=&\frac{1}{k}\vec \nabla \times\vec
A^{\,[L]}_m(M;k,\vec x\,)\,. 
\eeqa

The two-photon contribution to the polarizability is given by
\beqa
P_{fi}^{TPA,  J}(M^{\nu'}L',M^{\nu}L,k',k)&=&2\pi(-)^{L+J+1}\frac{\hat L\hat
  L'}{\hat J}\nonumber\\
&&\hspace*{-4.5cm}\langle I_f e_f||\int d^3x\, d^3y\,
\sum_{lm}\Big[A^{[L']}_l( M^{\nu'};k,\vec x\,){B}_{lm}(\vec
  x,\vec y\,) A^{[L]}_m(M^\nu;k,\vec y\,)\Big]^{[J]}||
I_i e_i\rangle\,.
\eeqa
The evaluation of the TPA contribution to the polarizabilities is
straightforward once the TPA operator ${B}_{lm}(\vec
x,\vec y\,)$ is given. 

For the resonance contribution, one finds by evaluating the reduced
matrix element in standard fashion (see e.g.~\cite{Edm57})
\beqa
P_{fi}^{RA, J}(M^{\nu'}L',M^{\nu} L,k',k)&=&2\pi(-)^{L+I_f+I_i}\hat L\hat L'
\nonumber\\&& 
\hspace*{-3.9cm}\times \sum\hspace{-.5cm}\int_{e_n, I_n}\left[
\left\{\begin{matrix}
L&L'& J \cr I_f &I_i& I_n \cr 
\end{matrix}\right\}\,
\frac{\langle I_f e_f||M^{\nu'[L']}(k')||I_n e_n\rangle
\langle I_n e_n||M^{\nu[L]}(k)||I_i e_i\rangle}
{e_n-e_i-k_0+\frac{\vec k\cdot(2\vec P_i+\vec k)}{2M_A}
-i\varepsilon}\right.
\nonumber\\&&\left.
\hspace*{-4.6cm}+(-)^{L+L'+J}
\left\{\begin{matrix}
L'&L& J \cr I_f &I_i& I_n \cr 
\end{matrix}\right\}\,
\frac{\langle I_f e_f||M^{\nu[L]}(k)||I_n e_n\rangle
\langle I_n e_n||M^{\nu'[L']}(k')||I_i e_i\rangle}{e_n-e_i+k_0'
-\frac{\vec k^{\,\prime}\cdot(2\vec P_i-\vec k^{\,\prime})}{2M_A}
-i\varepsilon}
\right].\label{pol-res}
\eeqa 
Obviously, the calculation of the resonance part is more involved
because of the summation over all possible intermediate states $|I_n
e_n\rangle$ with angular momentum $I_n$ and intrinsic energy $e_n$. 

The optical theorem in (\ref{opt-th}) allows one to relate the
imaginary part of the scalar polarizabilities of elastic scattering 
$P_{ii}^{ 0}(M^{\nu}L,M^{\nu}L,k,k)$ to the the partial contribution $\sigma(
M^{\nu}L) (k)$ of the multipole $M^{\nu}L$ to the total unpolarized photo 
absorption cross section $\bar\sigma_{tot}$ 
\beqa
\hspace*{-.5cm}\bar\sigma_{tot}(k)&=&
\sum_L\Big(\sigma(EL) (k)+\sigma(ML) (k)\Big) \nonumber\\
&=&\frac{4\pi}{2k(2I_i+1)}\,\sum_{\lambda M}
Im\,T^{ii}_{\lambda \lambda MM}(\vec k,\vec k)
\nonumber\\
&=&
\frac{4\pi}{k\widehat I_i}\sum_L \frac{(-)^{L+1}}{\widehat L}\,
Im\Big[P^{RA,0}_{ii}(EL,EL,k,k)+P^{RA,0}_{ii}(ML,ML,k,k)\Big].
\eeqa
From this relation follows 
\beq
Im[P^{0}_{ii}(M^\nu L,M^\nu L,k,k)]=\frac{k}{4\pi}(-)^{L+1}
\widehat I_i\widehat L \sigma(M^\nu L) (k)\,.\label{im-p0}
\eeq

An important property of the scattering amplitude is the low energy
theorem~\cite{Fri75} according to which up to terms linear in the photon
momentum $k$ the scattering amplitude is completely determined by
global properties like charge, mass and magnetic moment. This means
for the polarizabilities that in the limit $k=0$ only the scalar
E1-polarizability is nonvanishing, i.e. 
\beq
P^{ J}(E1,E1)|_{k=0}=-\delta_{J0}\,\widehat
I_i\,\sqrt{3}\,\frac{e^2Z^2}{M_A}\,,\label{low-energy} 
\eeq
with $I_i$ as ground state spin. It corresponds to the Thomson
scattering 
amplitude. Internal properties like static electric and magnetic
polarizabilities contribute in the next nonvanishing order ($k^2$)
only. An extension of this theorem by a more general low energy
expansion of the polarizabilities has been discussed in
Ref.~\refcite{ArW86} with the result that for electric multipole
transitions $EL$ with even $L$ the contributions to $P^{
  J}(EL,EL)|_{k=0}$ up to order $2L-2$ vanish.

\subsection{The elastic scattering cross section}

The elastic scattering cross section for unpolarized photons and
targets is given by  
\beq
\frac{d\sigma_{elastic}}{d\Omega}=
\frac{c(P_ {i}, k, k')}{2(2I_i+1)}
\sum_{\lambda,\lambda',M_i,M_f} |T^{ii}_{M_f M_i,\lambda'\lambda}(\vec
k^{\,\prime},\vec k)|^2\,, 
\eeq
with a kinematic factor for collinear intial momenta
\beq
c(P_ {i}, k, k')=\frac{E_i(k_0+E_i -k_0')}
{(\widehat k'\cdot (k+P_i))^2}\,, 
\eeq
where $\widehat k'=k'/k'_0$.
For elastic scattering in the c.m.\ frame ($\vec k+\vec P_i=0$ and
$k_0=k_0'$) one has $c_{c.m.}(P_ {i}, k, k')=(M_A^2+k^2)/W^2$ with
$W=k_0+E_i$ as invariant mass. 

In terms of the polarizabilities one finds for the unpolarized cross
section~\cite{ArG69} 
\beqa
\frac{d\sigma_{elastic}}{d\Omega}&=&
\frac{ c(P_ {i}, k, k')}{2I_i+1}
\sum_{L',L,K',K,J}\,\,\sum_{\nu',\nu,\bar\nu',\bar\nu}
P_{ii}^{ J}(M^{\nu'}L',M^{\nu}L,k',k) \nonumber\\
&&\hspace*{3cm}\times P_{ii}^{ J}(M^{\bar\nu'}K', M^{\bar\nu}K,k',k)^*
g_J^{\nu' L' \nu L; \bar\nu' K' \bar\nu K}(\theta)\,,
\eeqa
where the angular functions depend on the scattering angle $\theta$
only and are given by
\beqa
g_J^{\nu' L'  \nu L;  \bar\nu' K' \bar\nu K}(\theta)&=&
\frac{(-)^{ J}}{2}\widehat J^{\,2} (-)^{L+K+\nu'+\bar \nu'}\nonumber\\
&&\times \sum_{j}\widehat j^{\,2}
(1+(-)^{L+K+j+\nu+\bar\nu})
(1+(-)^{L'+K'+j+\nu'+\bar\nu'})\nonumber\\
&&\times
\left(\begin{matrix}
L'&K'& j \cr 1 & -1&0 \cr
\end{matrix}\right)
\left(\begin{matrix}
L&K& j \cr 1 & -1 &0 \cr
\end{matrix}\right)
\left\{\begin{matrix}
L&K& j \cr K' & L'&J \cr
\end{matrix}\right\} P_j(\cos\theta)\,,
\eeqa
with $ P_j(\cos\theta)$ denoting a Legendre polynomial. For pure $E1$
transitions this expression simplifys considerably and one obtains
with scalar, vector and tensor polarizabilities
\beq
\frac{d\sigma_{elastic} (E1)}{d\Omega}=
\frac{c(P_ {i}, k, k')}{(2I_i+1)}
\sum_{J=0}^2|P_{ii}^{ J}(E1,E1)|^2\,g_J^{E1}(\theta)\,,
\label{xsect}
\eeq
where in an abbreviated notation the angular functions are 
\beqa
g_0^{E1}(\theta)&=&\frac{1}{6}\,(1+\cos^2\theta)\,,\label{g0}\\
g_1^{E1}(\theta)&=&\frac{1}{4}\,(2+\sin^2\theta)\,,\\
g_2^{E1}(\theta)&=&\frac{1}{12}\,(13+\cos^2\theta)\,. \label{g2}
\eeqa
In this case the angular distribution is symmetrical around 90$^\circ$
in the c.m.\ system. 

The generalization to polarized photons and oriented nuclei is presented
in Refs.~\refcite{ArG69} and \refcite{ArG66}.

\section{Applications}

Now I will discuss several applications: (i) Heavy deformed 
nuclei within the dynamic collective model of Danos and Greiner, (ii)
Photon scattering off $^{12}$C for energies in the region of the first
nucleon resonance, the $\Delta(1232)$, and finally (iii) photon scattering
off the deuteron as a means to study subnuclear degrees of freedom, 
for example meson exchange currents or the static polarizabilities of the neutron. 

\subsection{Photon scattering off complex nuclei in the giant
 dipole resonance region}

The giant dipole resonance (GDR) in medium and heavy weight nuclei can
well be explained as a collective 
phenomenon in the framework of the hydrodynamical model of Steinwedel
and Jensen as an oscillation of a proton fluid against a neutron  
fluid. According to this model, as pointed out independently by Danos
and Okamoto~\cite{DaO58}, the GDR will be split into two peaks
for an axially symmetric deformed nucleus  corresponding to different
frequencies for oscillations along and perpendicular to the symmetry
axis.  

In view of the additional collective surface degrees of freedom, Danos
and Greiner~\cite{DaG64} proposed in 1964 a unified dynamic
collective model of the giant resonances (DCM) which includes 
the coupling between the rotation-vibration surface degrees of freedom
and the giant
resonance d.o.f.\ leading to additional dynamic effects.  
A weak point of this approach is that it provides energies and
strengths of the GDR states at discrete energies only, but not the shape
nor the widths of the states. Usually a Lorentzian shape is assumed
with some simple adhoc model for the width, the parameters of which
are used for a fit procedure to the absorption cross section.

\begin{figure}[ht]
\centerline{\includegraphics[width=12cm]{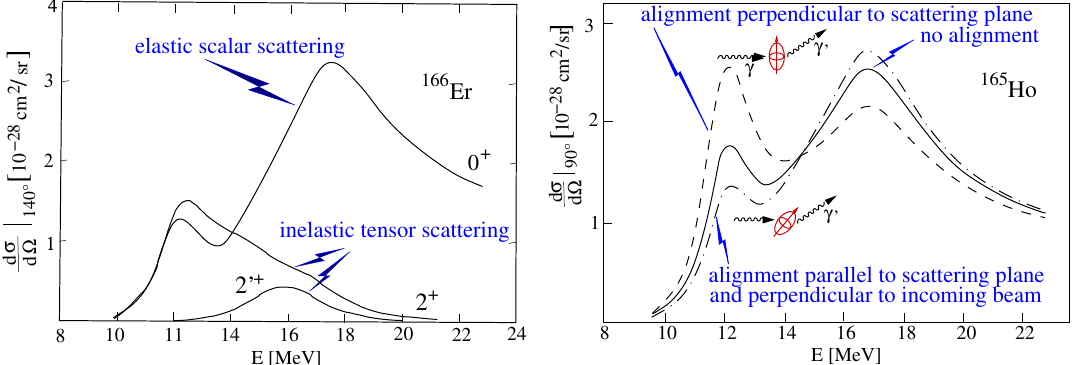}}
\caption{
Left panel: Calculated elastic and inelastic photon scattering cross 
sections at $140^\circ$ for $^{166}$Er (from Ref.~\refcite{ADG67}). 
Right panel: Elastic photon scattering cross sections for $^{165}$Ho 
(from Ref.~\refcite{ArG66}):
unoriented target: solid curve; aligned target:
(a) perpendicular to scattering plane: dashed curve,
(b) parallel to scattering plane and perpendicular to incoming
photon beam: dash-dot curve.}
\label{fig2}
\end{figure}

As a result of this dynamic coupling considerably strong dipole
transition strengths from the GDR states to the low lying rotational
and vibrational states appear leading to sizeable Raman scalar and
tensor scattering into these low lying collective states (see
Fig.~\ref{fig2}, left panel). Indeed, such Raman scattering has been
measured for $^{238}$U, $^{232}$Th and $^{209}$Bi by Jackson and
Wetzel~\cite{JaW72} although these authors found a significantly
weaker Raman cross section (about 40 \%) 
than predicted by the DCM. A similar reduction of the inelastic DCM
strenghts was found later for a series of vibrational medium weight
nuclei by Bowles et al.~\cite{BHJ81} A possible explanation of this
reduction could be a 10-15~\% nonresonant contribution of direct
transitions into the continuum which would appear in the scalar but
not in the tensor polarizability as has been discussed in
Ref.~\refcite{Are72}. 
 
Another interesting feature of the DCM is the fact that a deformed
nucleus with a nonvanishing ground state spin becomes dynamically
triaxial and thus optically
anisotropic (nonvanishing elastic tensor polarization) and thus its
absorption and scattering cross sections depend on the
nuclear orientation. The reason for this feature is the fact that for
a nonvanishing ground state spin $I\ge 1$ there is a tensor
contribution to elastic scattering which, however, does not show up
for an unoriented nucleus. This is shown in the right panel of
Fig.~\ref{fig2} for $^{165}$Ho having a ground state spin $I=7/2$.
Experimentally such a dependence of the photon absorption cross
section of $^{165}$Ho on the nuclear orientation was found by Ambler
et al.~\cite{AFM65} . 

\begin{figure}[ht]
\centerline{
     {\includegraphics[width=4.32cm]{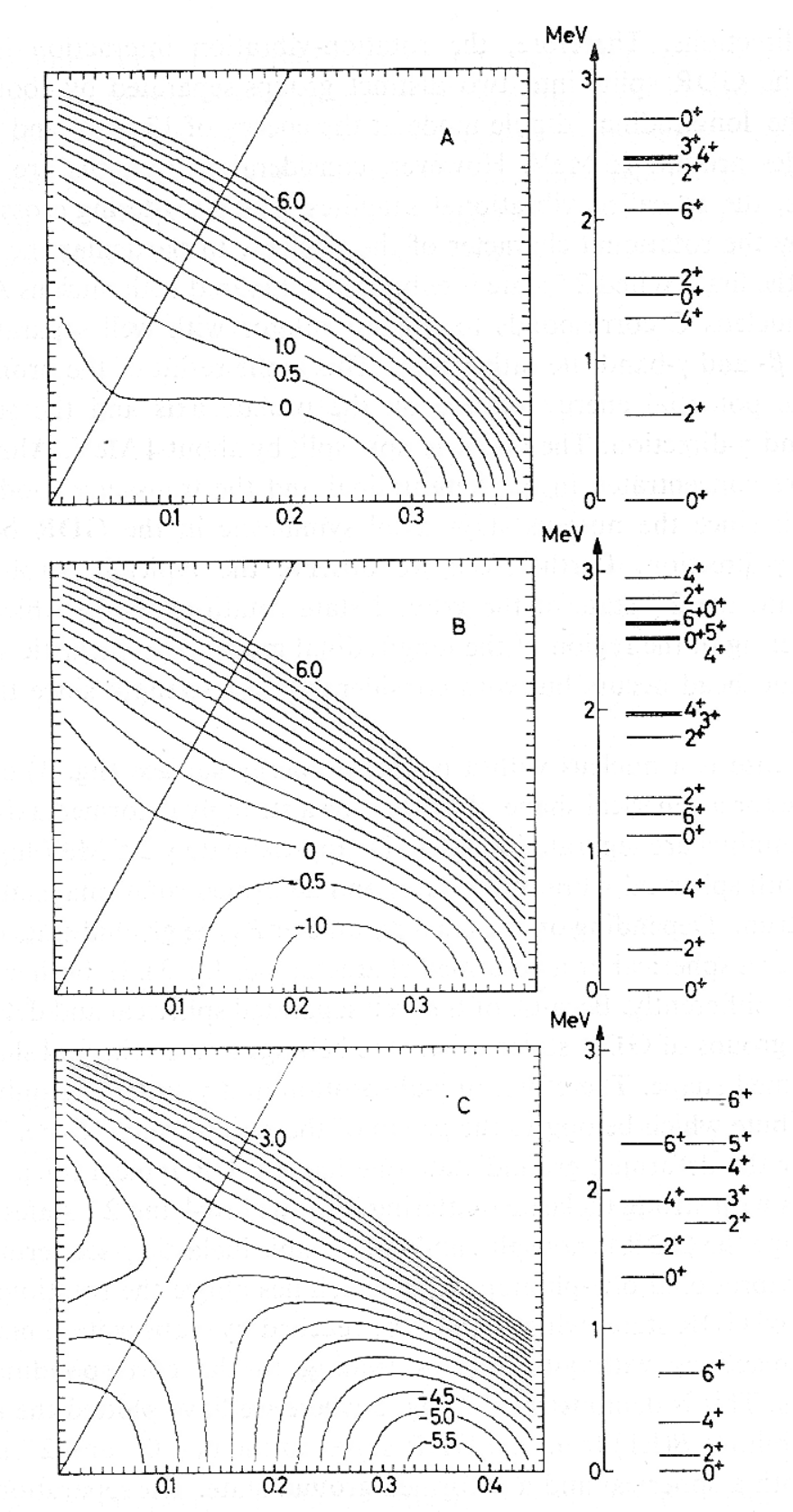}}
     {\includegraphics[width=7.2cm]{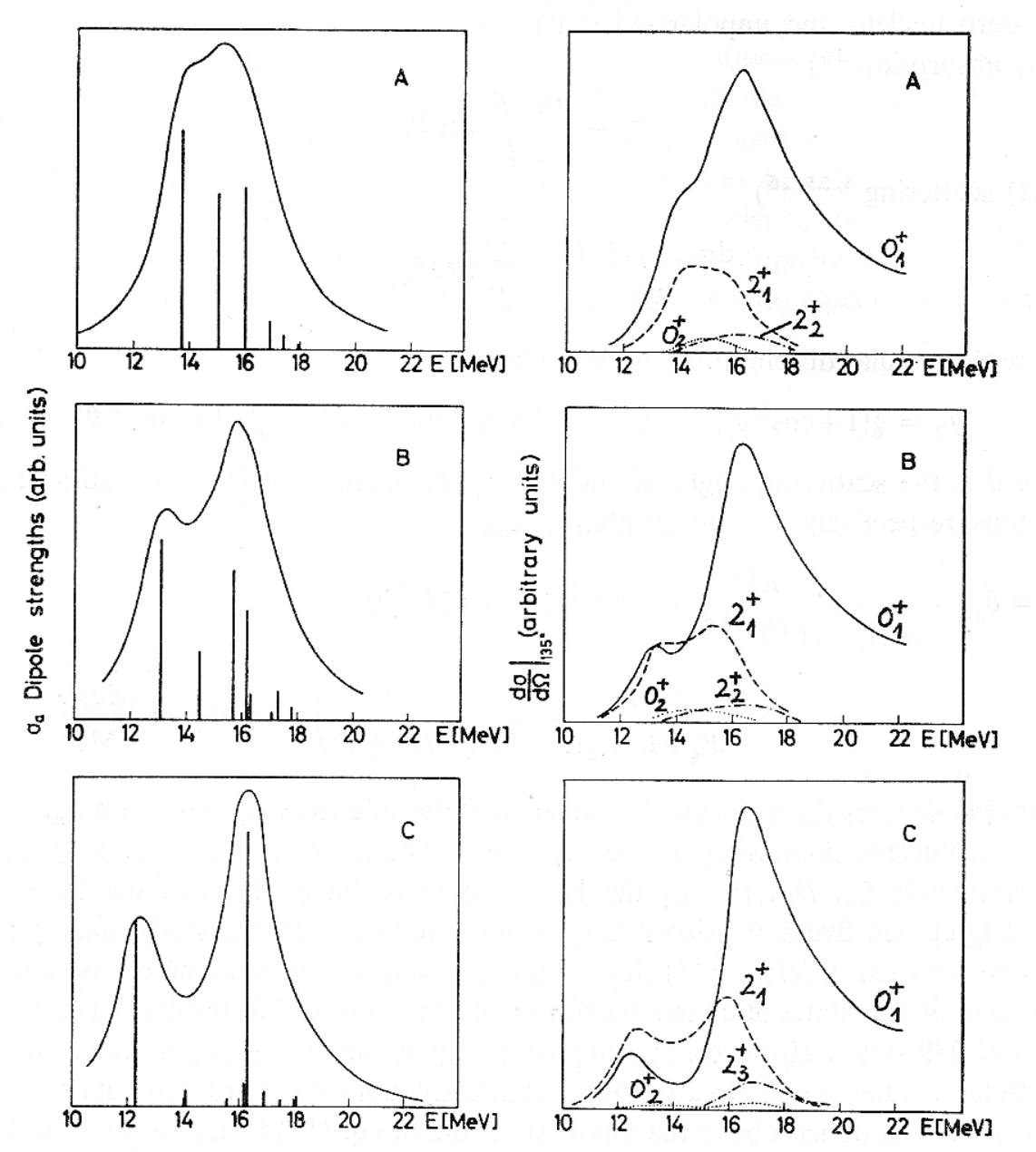}}}
\caption{Transition study from a vibrational (A) to a strongly deformed
  nucleus (C) (from Ref.~\refcite{RGA70}):
Left panels: Collective potential energy surfaces and low-energy
spectra. Middle panels: Dipole strengths and $\gamma$-absorption cross
sections. Right panels: Elastic and Raman $\gamma$-scattering cross
sections to low lying collective states.}
\label{fig4}
\end{figure}

Subsequently, the DCM was further developed in order to describe a
much more  
general class of potential energy surfaces for the low energy collective 
d.o.f.\ allowing a unified description of the GDR for nuclei with quite 
different collective characteristics by Rezwani et
al.~\cite{RGA70,RGA72}. An example,  taken from Ref.~\refcite{RGA70},
is shown in  
Fig.~\ref{fig4} displaying the collective potential energy surface,
the absorption strength and cross section and the elastic and
inelastic scattering cross sections for three different types of nuclei,
resembling a transition 
from a nucleus with an anharmonic vibrational character (A) to a
strongly deformed rotational nucleus (C). 

Case A represents a vibrational nucleus with strong anharmonicities,
resulting in a shift and splitting of the surface two-phonon
triplet. Furthermore, a slight axially symmetric deformation leads to
a nonvanishing intrinsic quadrupole moment. Since the nucleus is easily
deformable in $\beta$- and $\gamma$-directions, it appears dynamically
triaxial resulting in a slight splitting of the dipole strength into three
equally spaced states. Moreover, considerable inelastic tensor
$\gamma$-scattering into the low-lying $2^+$-states is observed. 

The opposite situation of a strongly deformed nucleus is displayed by
case C. It corresponds to a good axially symmetric rotator with
clearly separated ground, $\beta$- and $\gamma$-rotational
states. Accordingly the GDR is split into two distinct peaks, and
strong Raman-scattering via the tensor polarizability into the
$2^+$-state of the ground state rotational band appears. The
transition between these two extreme cases is represented by case B. 

\subsection{Photon scattering off complex nuclei in the 
$\Delta(1232)$  region}

The influence of internal nucleon degrees of freedom constitutes an
important field of research in nuclear and medium energy physics. In
particular, 
the role of the lowest excited state of the nucleon, the $\Delta(1232)$
resonance, has been studied in the low energy
domain~\cite{WeA78} as well as 
in the energy region of real $\Delta$ excitation in pion
production and photo absorption. 

For photon scattering in the energy region of about 300 MeV the
simplest model is a static approach assuming pure $M1$-scattering off
the individual nucleons, whereas the nuclear structure is manifest
only via the elastic form factor with respect to the momentum
transfer. Then the scattering matrix has the simple form~\cite{AWR85} 
\beq
T^{fi}_{\lambda'\lambda} = \langle f| \sum_{l=1}^A t_{\lambda'\lambda}
(l)e^{-i(\vec k-\vec k^{\,\prime})\cdot \vec r_l}|i\rangle\,,
\eeq
where the elementary scattering operator is given by
\beqa
t_{\lambda'\lambda}(l) &=& c^2 
\,\Big(\frac{\vec e_{\lambda'}^{\,\prime *}\cdot
[\vec \sigma_{N\Delta}^l\times \vec k^{\,\prime}]\,
\vec e_{\lambda}\cdot
[\vec \sigma_{\Delta N}^l\times \vec k\,]}{M_\Delta - M_N - k_0
-i\Gamma_\Delta/2}
\nonumber\\&&
+\frac{\vec e_{\lambda}\cdot
[\vec \sigma_{N\Delta}^l\times \vec k\,]\,
\vec e_{\lambda'}^{\,\prime *}\cdot
[\vec \sigma_{\Delta N}^l\times \vec k^{\,\prime}]}{M_\Delta - M_N + k'_0
-i\Gamma_\Delta/2}\,\Big)\,.
\eeqa
The $N\to \Delta$ spin transition matrix $\vec \sigma_{N\Delta}^l$ is defined as in
Ref.~\refcite{WeA78}. Furthermore, 
\beq
c=G_{M1}^{\Delta N}(M_\Delta+M_N)/(4 M_\Delta M_N)\,,
\eeq
where $M_N$, $M_\Delta$ and $\Gamma_\Delta$ denote respectively 
nucleon and $\Delta$ mass and width. The latter and the magnetic
transition strength $G_{M1}^{\Delta N}$ are fit to the experimental
photo absorption cross section of the nucleon in the $\Delta$
resonance region.  

\begin{figure}[ht]
\centerline{
  \subfigure[$^{12}$C]
     {\includegraphics[width=6.1cm]{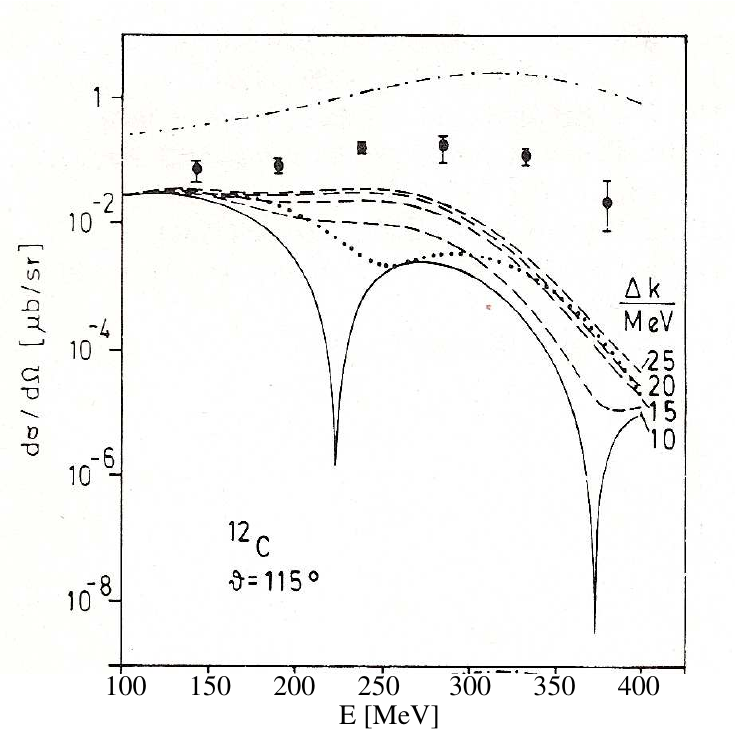}\label{Fig-C}}
  \hspace*{.05cm}
  \subfigure[$^{208}$Pb]
     {\includegraphics[width=6.2cm]{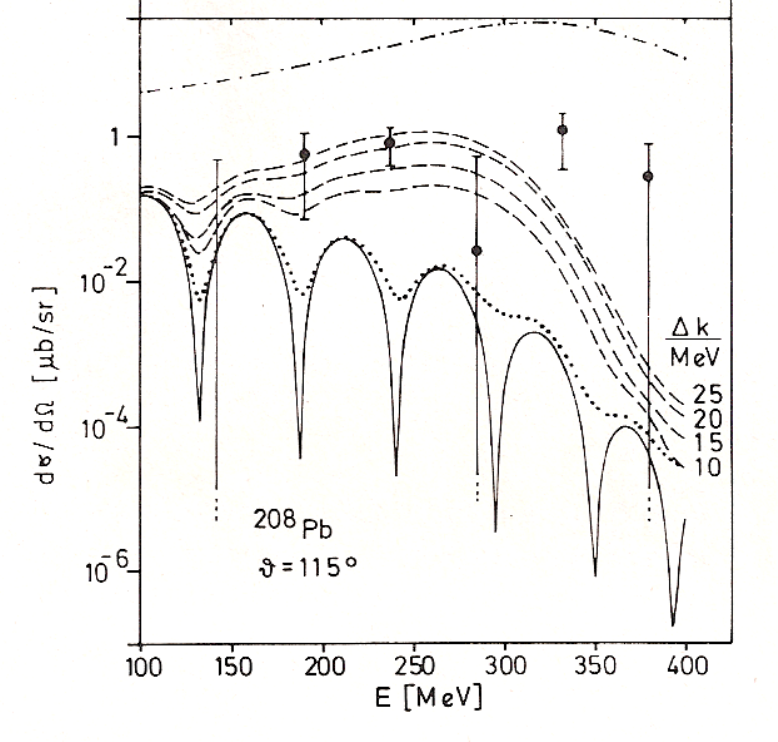}\label{Fig-PB}}}
\caption{Elastic and quasielastic photon scattering cross sections at
 $\theta=115^\circ$ for $^{12}$C and $^{208}$Pb (from
 Ref.~\refcite{AWR85}). Experimental data 
 from Ref.~\refcite{HaZ84}. Solid curve: elastic scattering in static
 approach; dashed curves include inelastic contributions up to an
 excitation energy $\Delta k$; dotted curve: sum rule approach (see
 Ref.~\refcite{AWR85} for further details); dash-dotted curve: incoherent
 sum of elastic  elementary $\gamma-N$ scattering cross
 section.} \label{Fig-C-PB} 
\end{figure}

In this simple approach nuclear structure enters only via the nuclear
form factors of mass and spin (for more details see
Ref.~\refcite{AWR85}). A comparison of this approach with experiment
for $^{12}$C and $^{208}$Pb is shown in Fig.~\ref{Fig-C-PB}.  

It turns out, that the resulting calculated cross section largely
underestimates 
the experiment by orders of magnitude for both nuclei. Collective
effects in a more refined $\Delta$-hole model for $^{12}$C by Koch et
al.~\cite{KMO84} lead to a slight enhancement but the discrepancy
remained essentially. However, Hayward and Ziegler~\cite{HaZ84}
already had pointed out that the finite energy resolution with respect
to the measured scattered
photons of about 10\,\% would lead to the inclusion of corresponding
inelastic contributions in the experimental data. 

Subsequently, such inelastic
contributions have been studied in Ref.~\refcite{AWR85} within two
simple models: (i) Inelastic scattering into 1-particle-1-hole
excitations with excitation 
energies within an energy interval $\Delta k$ , and (ii) a sum rule
approach to the giant resonances as final states (see
Ref.~\refcite{AWR85} for further details). The results of these two
models are also shown in Fig.~\ref{Fig-C-PB}. Both approaches show an
enhancement due to inelastic contributions, particularly strong for
the 1p -1h approach. However, for $^{12}$C still sizeable strength is
missing and a more thorough theoretical treatment is needed. 

\subsection{Photon scattering off the deuteron}

Now I would like to turn my attention to more fundamental studies with
respect to the role of subnuclear degrees of freedom like meson
exchange currents (MEC) and and internal nucleon degrees of freedom in
terms of nucleon resonances (isobar currents IC) in electromagnetic
reactions. Lightest nuclei 
present ideal laboratories for such studies and I will concentrate on
the deuteron for which extensive studies on electromagnetic reactions
exist~\cite{Are84,Are86,ArS91,ALT05}. 

\begin{figure}
\centerline{\includegraphics[width=9cm]{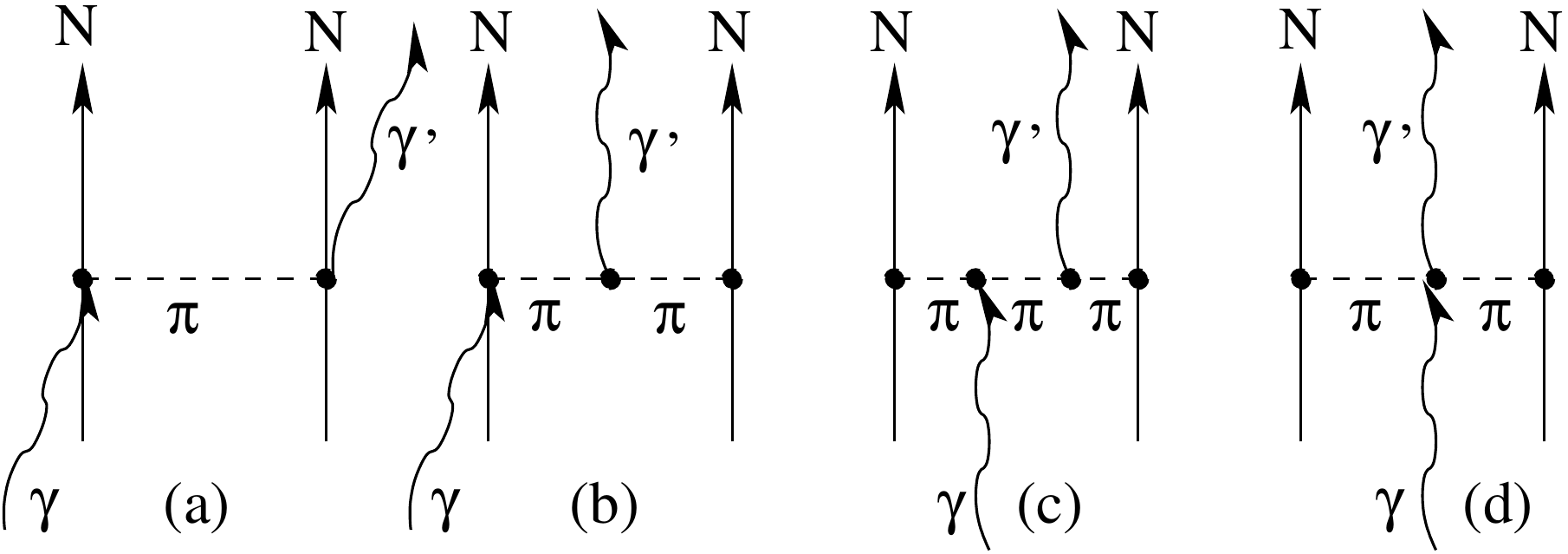}} 
\caption{Diagrammatic representation of pion exchange TPA
  contributions to the  two-photon amplitude.} 
\label{Fig-TPA-MEC}
\end{figure}

In photon scattering meson exchange effects do not only appear via MEC
in the resonance amplitude but also as additional 
contributions in the TPA as derived in~\cite{Are80} for a $NN$ one-pion 
exchange potential (see Fig.~\ref{Fig-TPA-MEC} for the corresponding
diagrams) as well as for isobar contributions. 

\begin{figure}[ht]
\centerline{
\includegraphics[width=6.cm]{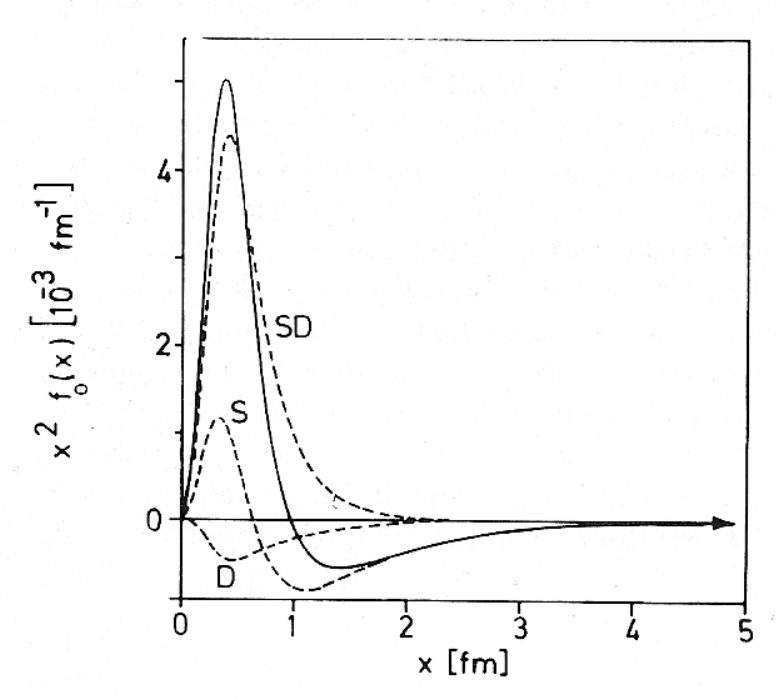}
  \hspace*{.2cm}
\includegraphics[width=5.5cm]{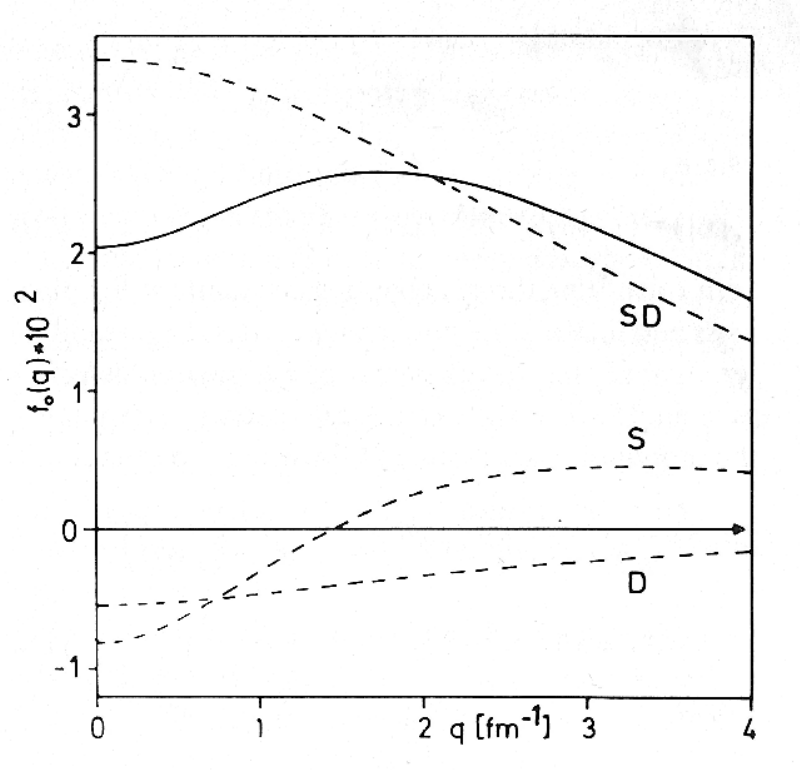}
}
\caption{Left panel: Monopole part of $\pi$-exchange transition density for
  the deuteron (solid curve) from the $\pi$-TPA diagram (d) of
  Fig.~\ref{Fig-TPA-MEC}. Dashed curves show separate 
  contributions from $S$- and $D$-states and $S-D$ interference.
Right panel: Form factor of transition density displayed in the left
panel, again with separate contributions (from Ref.~\refcite{WeA84}).}  
\label{Fig-pion-tra-dens} 
\end{figure}

It is intriguing, and indeed had been suggested, that diagram (d) of
Fig.~\ref{Fig-TPA-MEC} 
allows the extraction of a density form factor of charged mesons in
the nucleus (see references in Ref.~\refcite{WeA84}). However, a
careful analysis has shown that such an interpretation is not
possible~\cite{WeA84}. It turns out that the TPA of an exchanged pion 
(diagram (d) of Fig.~\ref{Fig-TPA-MEC}) reflects the form factor of a
pion transition density between the two nucleons and not the virtual
pion density inside the deuteron. This feature is illustrated in
Fig.~\ref{Fig-pion-tra-dens}, 
where the monopole part of this transition density is displayed.
Moreover, as already mentioned above, the diagrams in 
Fig.~\ref{Fig-TPA-MEC} are not by itself gauge invariant and thus not
separately measurable. 

A first realistic calculation of elastic deuteron photon scattering
with inclusion of subnuclear effects 
has been carried out in Ref.~\refcite{WeA83} taking as reference frame
the photon deuteron Breit frame, then $k'=k$. Furthermore, small
contributions from the c.m.\ motion have been neglected. For
unpolarized photons and deuterons only the scalar polarizability
$P^{0}_{ii}(M^{\nu}L,M^{\nu}L,k,k)$ contributes to the elastic
scattering cross section. The calculation of the resonance 
amplitude requires the summation over all excited energies and
all possible intermediate states. 

As detailed in Ref.~\refcite{WeA83}, a subtracted dispersion
relation for the individual polarizabilities of given multipolarity
has been assumed using Eq.~(\ref{im-p0}) for the imaginary part of the
resonance amplitude (the TPA is real). Thus the real part of the
scalar polarizability is determined from
\beqa
Re (P^{RA,0}_{ii}(M^{\nu} L,M^{\nu} L,k,k))&-&\delta_{\nu
  0}\delta_{L1}\,P^{RA,0}_{ii}(E1,E1,0,0)\nonumber\\
&=&(-)^{L+1}\widehat L\widehat I_i\frac{k^2}{2\pi^2} 
P\int_{k_{th}}^\infty dk'\frac{\sigma(M^{\nu} L)(k')}{k^{\prime 2}-k^2}\,.
\label{dispersion}
\eeqa
Here, $k_{th}$ denotes the threshold energy for photo
absorption. Since the integral extends in principle up
to infinity, one has to include also contributions from particle
production above the corresponding threshold energy, e.g.\ from the
total photo pion production cross section for energies above the pion
production threshold. These were neglected limiting this approach to
the low energy region. For the evaluation of the partial contributions 
$\sigma(M^{\nu} L)(k)$ to photo disintegration the Reid soft core
potential had been used. Besides the one body current and TPA
$\pi$-meson exchange currents also $N\Delta$ and $\Delta\Delta$ isobar
configurations have been included. 

\begin{figure}
\centerline{\includegraphics[width=13.5cm]{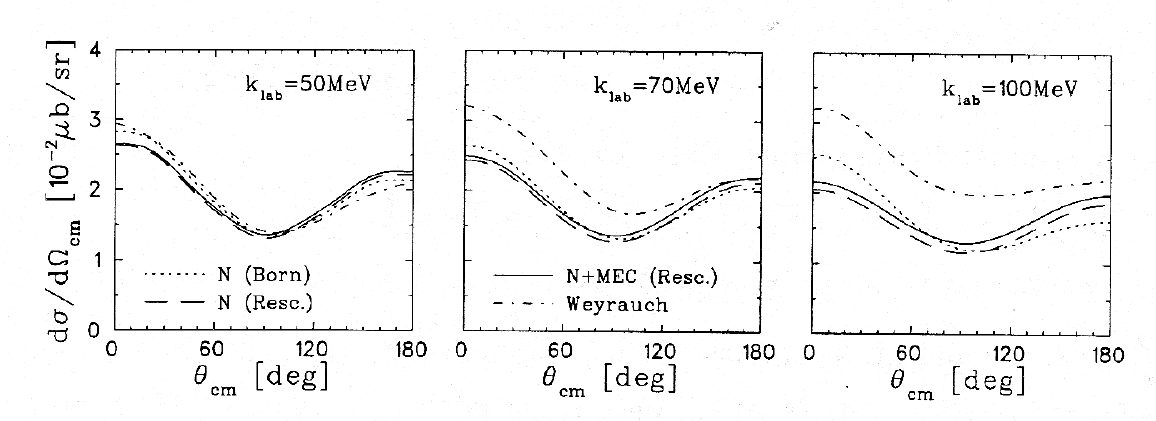}} 
\caption{Differential scattering cross sections for photon energies
  50, 70 and 100 MeV from Ref.~\refcite{WWA95}. Notation of curves:
  dotted (labeled ``N (Born)'') without FSI and MEC 
  beyond Siegert, dashed (``N (Resc.)'') including FSI but no MEC,
  and full (``N+MEC (Resc.)'') with FSI 
  and MEC. Dash-dot curves (``Weyrauch'') represent a calculation with
  a separable   interaction from Ref.~\refcite{Wey90}.}
\label{Fig-photon-deuteron}
\end{figure}

A more recent realistic calculation from my group by Wilbois et
al.~\cite{WWA95} has been based on the solution of the off-shell
NN-scattering matrix instead of using dispersion relations. Again
inelasticities, which appear above pion production threshold, are
excluded, limiting this approach also to low photon energies. For the 
deuteron bound state and the final state interaction (FSI) the
realistic Bonn OBEPQ-B potential has been used. MEC contributions from 
pion- and rho-meson exchange to the current and the TPA amplitude have 
been included. For the electric transitions so-called Siegert
operators have been used, incorporating thus implicitly the major part
of MEC contributions~\cite{Are81}. Fig.~\ref{Fig-photon-deuteron}
displays the resulting differential scattering cross section at three
photon energies, 50, 70, and 100 MeV. For comparison the results of
another approach by Weyrauch~\cite{Wey90} using a separable
NN-interaction are also shown. Contributions from
internal nucleon structure as manifest, e.g.\ in nucleon
polarizabilities, were neglected in both calculations. 

Comparing the dotted with the dashed
curves one notes a sizeable increase from FSI with increasing energy,
at forward  angles a reduction and in the backward region an increase,
thus reducing the strong asymmetry of the case without FSI
considerably. The additional MEC effects
beyond the Siegert operators are quite small for the lowest energy,
but become more 
pronounced at 100 MeV, leading to an overall increase of the cross
section. The increasing difference to Weyrauch's result~\cite{Wey90}
indicate that the separable interaction used in 
that work is not appropriate at higher energies. 

\begin{figure}
\centerline{\includegraphics[width=11cm]{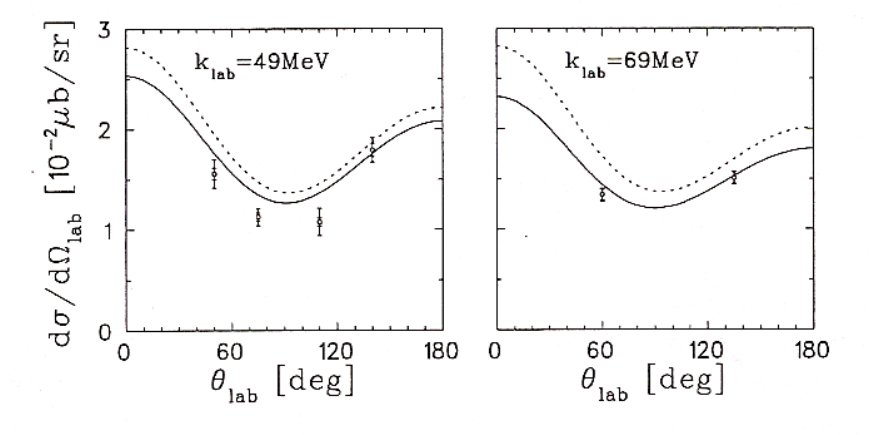}} 
\caption{Differential scattering cross sections for photon energies of
  49 and 69 MeV (from Ref.~\refcite{WWA95}). Experimental data from
Ref~\refcite{Luc94}. Solid curves: 
complete calculation with nucleon polarizabilities included. Dashed
curves: without nucleon polarizabilities.}
\label{Fig-photon-deuteron-pol}
\end{figure}

A comparison with experimental data is shown in
Fig.~\ref{Fig-photon-deuteron-pol}. One readily notes a systematic
overestimation of the experimental results (dashed curves). An improved
description is achieved if the influence of the internal nucleon
structure in terms of nucleon polarizabilities is considered (full
curves). The additional introduction of the free neutron and proton
polarizabilities reduces the cross sections sizeably and gives a better
description of the experiment. In view of the fact, that the neutron
polarizabilities are not directly measurable, one turns the argument
around in order to determine them from photon scattering off the
deuteron. However, this procedure is not completely free from model
dependencies. 

Since then quite a few more theoretical investigations of this
reaction using various approaches have been
published~\cite{GGD}, where the major
emphasis had been laid on the extraction of the neutron
polarizabilities (for recent results see Ref.~\refcite{Mye15}). 

In fact, the deuteron is often used to determine internal
neutron properties like polarizabilities and electromagnetic form
factors, which otherwise are not available, since free neutron targets
are absent. The question of off-shell modification of such internal
nucleon properties is thereby left open with the hope that such
effects are small since the deuteron is quite a loosly bound system. 

\begin{figure}
\centerline{\includegraphics[width=7.5cm]{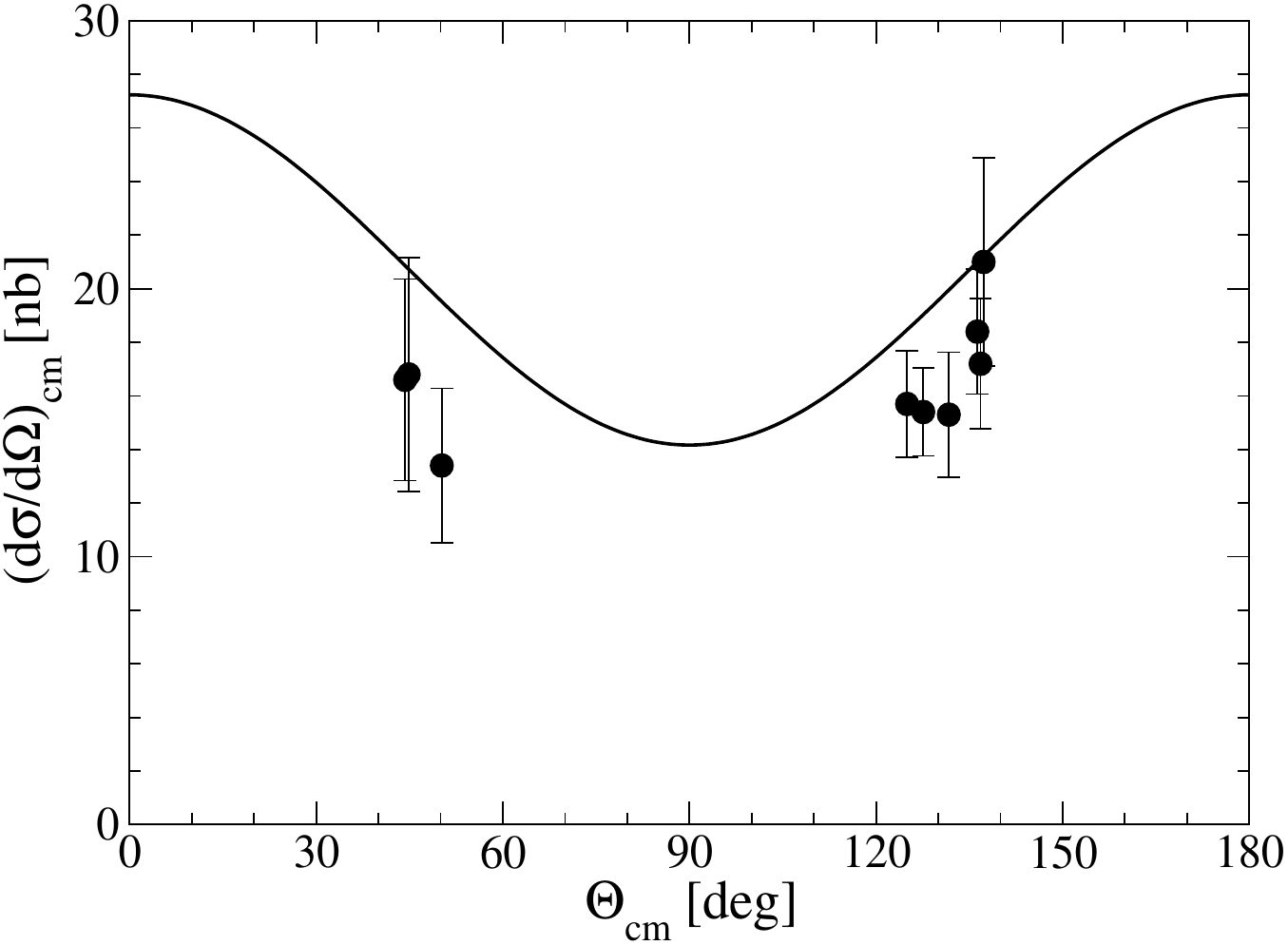}} 
\caption{Differential scattering cross section in the unretarded
  dipole approximation calculated with the Lorentz integral transform
  method for a photon energy of 55 MeV (from
  Ref.~\refcite{BLA11}). Experimental data from Ref.~\refcite{Lun03}.}
\label{Fig-Lit}
\end{figure}

Another interesting aspect which reactions on the deuteron offer
is that the deuteron can also serve as a test ground for new theoretical
methods. As an illustrative example, I will consider recent work
based on the Lorentz integral transform (LIT)  method (for a review of
this method see
Ref.~\refcite{ELO07}). The LIT method is particularly suited for
theoretical studies of reactions on complex nuclei. It allows the
calculation of reactions without the need of determining the complex
final scattering states, reducing this problem to the solution of a
bound state 
equation. Therefore, as a test case for the application of the LIT
method to photon scattering reactions, the LIT has
been applied to the deuteron for a pure unretarded $E1$
radiation by Bampa et al.~\cite{BLA11}.  

The resulting cross section at one energy is
shown in Fig.~\ref{Fig-Lit} together with recent experimental
data. Nucleon polarizabilities were not included. The
slight difference to the more complete calculation at 50 MeV in
Fig.~\ref {Fig-photon-deuteron} is caused by the restriction to $E1$
radiation which is also the reason for the completely symmetric
angular distribution around 90$^\circ$ (see Eq.~(\ref{g0})) in
contrast to the slight asymmetry at 50 MeV in
Fig.~\ref{Fig-photon-deuteron}.  

\section{Conclusions}

With these few examples, I hope to have convinced the reader, that
nuclear photon scattering reactions has been and still is a wide and
interesting field of 
research in nuclear structure studies from low energy collective
properties to the influence of subnuclear degrees of freedom as
manifest e.g.\ in meson exchange and isobar currents. Hopefully, in
the future more experimental results will be available, although the
cross sections are quite small compared to hadronic reactions. Of
particular interest are scattering experiments on complex nuclei at
higher energies in the region of the first nucleon resonance, in order
to study the behavior of a nucleon resonance in a nuclear
environment.

\end{document}